\documentclass[aps, pre, reprint, amsmath, amssymb, superscriptaddress, floatfix]{revtex4-1}

\usepackage{graphicx} 
\usepackage[caption=false]{subfig}  % use for side-by-side figures
\renewcommand{\vec}{\mathbf}

\begin{document}

\title{Order-disorder transition in repulsive self-propelled particle systems}
\author{Takayuki Hiraoka}
\affiliation{Department of Applied Physics, The University of Tokyo, Tokyo 113-8656, Japan}
\author{Takashi Shimada}
\affiliation{Department of Applied Physics, The University of Tokyo, Tokyo 113-8656, Japan}
\author{Nobuyasu Ito}
\affiliation{Department of Applied Physics, The University of Tokyo, Tokyo 113-8656, Japan}
\affiliation{RIKEN Advanced Institute for Computational Science, Kobe 650-0047, Japan}

\begin{abstract}
We study the collective dynamics of repulsive self-propelled particles. The particles are governed by coupled equations of motion that include polar self-propulsion, damping of velocity and of polarity, repulsive particle-particle interaction, and deterministic dynamics. Particle dynamics simulations show that the collective coherent motion with large density fluctuations spontaneously emerges from a disordered, isotropic state. In the parameter region where the rotational damping of polarity is strong, the systems undergoes an abrupt shift to the absorbing ordered state after a waiting period in the metastable disordered state. In order to obtain a simple understanding of the mechanism underlying the collective behavior, we analyze binary particle scattering process. We show that this approach correctly predicts the order-disorder transition at dilute limit. The same approach is expanded for finite densities, although it disagrees with the result from many-particle simulations due to many-body correlations and density fluctuations.
\end{abstract}

\maketitle

\section{Introduction}
Following the seminal works by Vicsek et al. \cite{Vicsek1995} and Toner and Tu \cite{Toner1995, Toner1998}, the interest of physicists in collective motion of self-propelled particle (SPP) systems has been growing in last two decades. The main aim of the field is to obtain an unified understanding over non-trivial ordering behaviors seen in a group of living organisms in various length scales, from flocks of birds and schools of fish \cite{Ballerini2008a, Cavagna2010, Katz2011, Bialek2012, Gautrais2012, Lopez2012} to tissue cell migration in monolayers \cite{Szabo2006, Poujade2007, Szabo2010, Anon2012, Serra-Picamal2012} and bacterial colonies \cite{Zhang2010, Peruani2012, Wensink2012} and even to the intracellular structure such as actin filaments and microtubules \cite{Schaller2010, Sumino2012}.
A well established approach in the field is the Vicsek-type SPP model \cite{Vicsek1995, Gregoire2004,Chate2008}. It assumes the particles drive themselves with a constant speed while adjusting their direction of motion parallel to their neighbors' velocities. It has been shown that the non-equilibrium character of the systems leads to development of long-range order and giant density fluctuations, which are unusual in equilibrium systems \cite{Toner1995, Toner1998, Simha2002, Ramaswamy2003, Chate2006}.

Experiments using vibrated grains and driven colloids \cite{Narayan2007, Aranson2008, Deseigne2010, Deseigne2012, Buttinoni2013} suggest that the similar properties can also be found in non-living systems. However, the Vicsek model is not likely to illustrate the microscopic nature of these systems, where particles do not ``compute" the average of their neighbors' velocities. It is rather natural to assume that the elements interact with repulsion through excluded volume. Pedestrian movement, another major target of SPP studies, is also known to be well described by models that assume short-ranged repulsive force from neighboring people \cite{Helbing1995, Helbing2000}. Therefore, it is of great importance to capture dynamics of SPP systems where repulsive interactions are dominant. 

Previously, Hanke et al. \cite{Hanke2013} proposed a model of soft, repulsive active particles, explored its collective behavior over parameters, and found that a polarized state emerge at certain parameter region. However, their model has some inconsistencies at microscopic level when compared to actual granular or colloidal matter. First, the particles are not always as soft and strongly overdamped as the model requires them to be in order to achieve the collective motion. Second, the origin of the noise is unclear, since thermal fluctuation is not relevant in the length scale we deal with. 

Here we consider a simple model of repulsive SPP that can be applied to various systems from motile cells, active colloids and grains to pedestrian crowds. This generalized model obeys deterministic equations of motion and interacts with other particles by short-range repulsion. We explore the phase diagram and discuss the nature of the phase transition. In section 4 and 5, we analyze the emergence of collective motion from binary scattering process.

\begin{figure*}[tb]
\subfloat[$t = 3600$] {
	\centering
	\includegraphics[width=0.43\hsize]{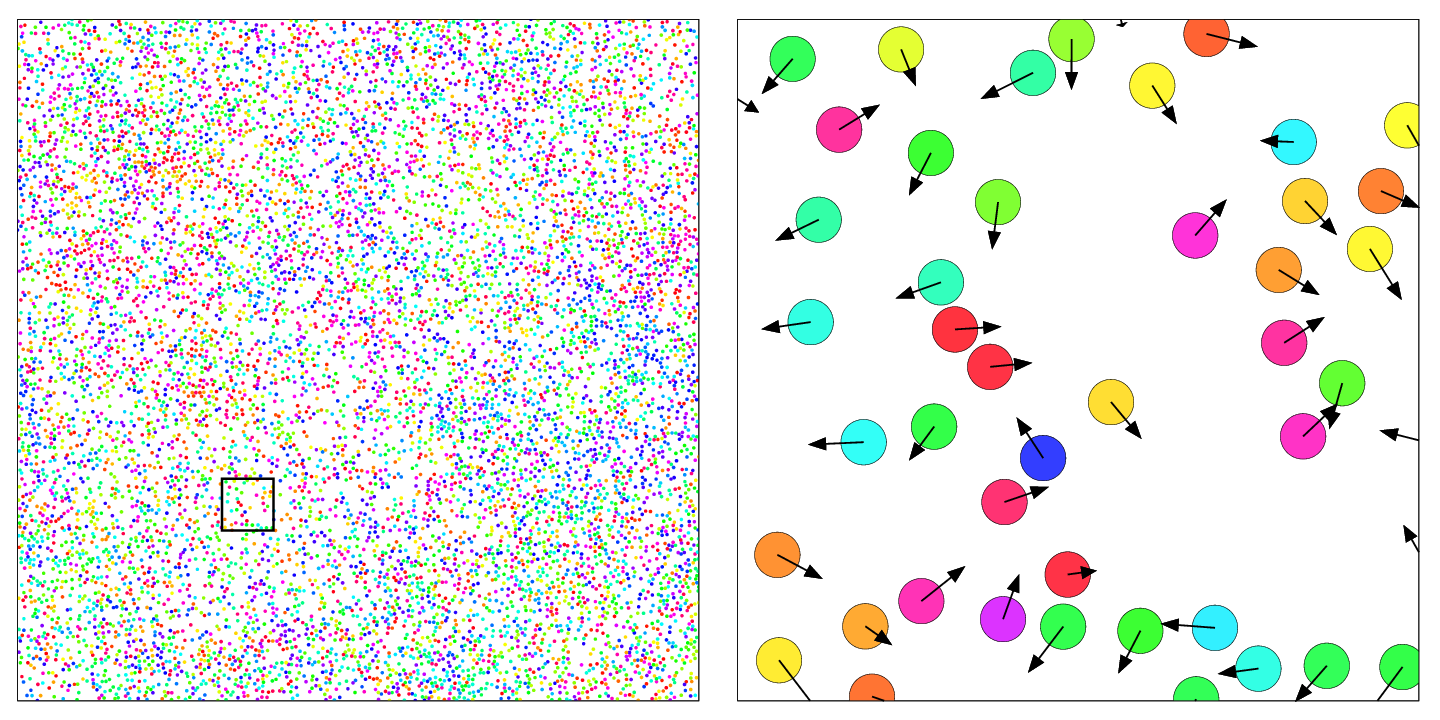}
	\label{fig:3600}
}
\subfloat[$t = 3980$] {
	\centering
	\includegraphics[width=0.43\hsize]{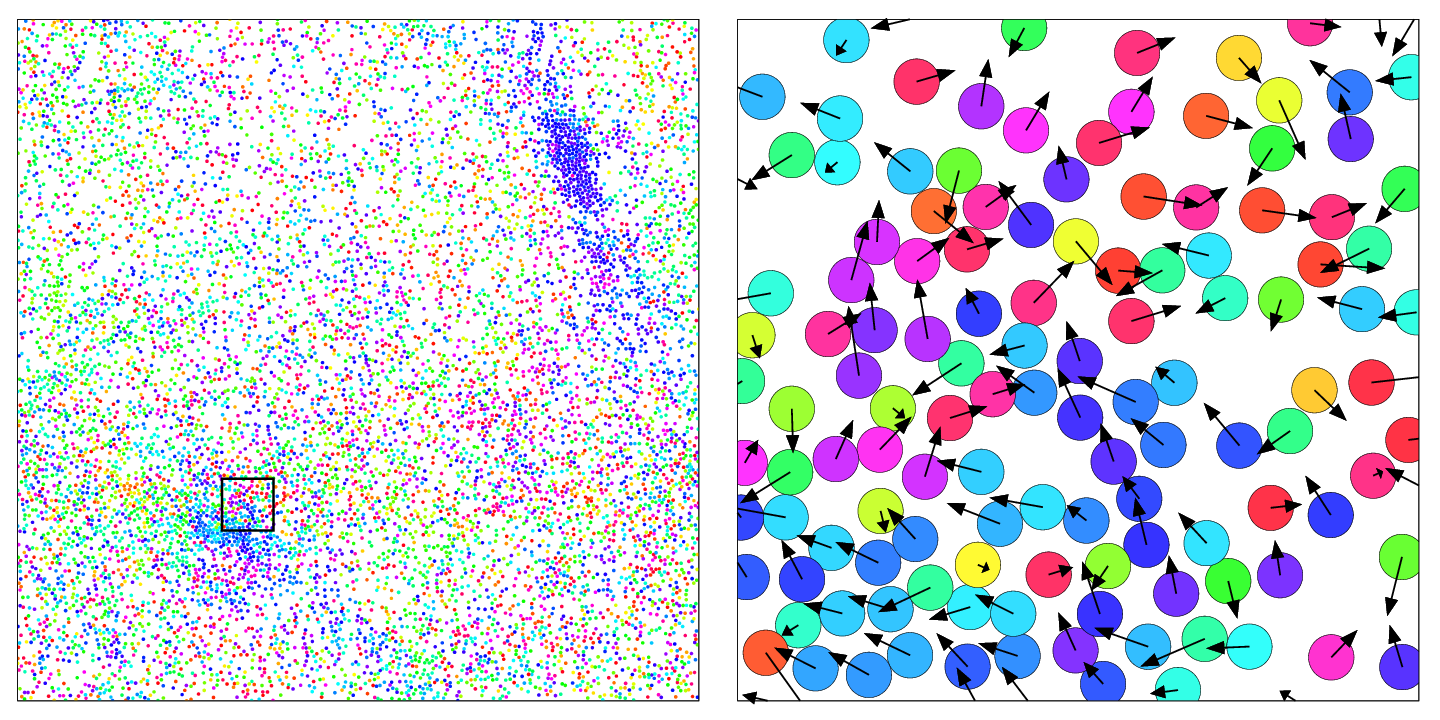}
	\label{fig:3980}
} \\
\subfloat[$t = 4120$] {
	\centering
	\includegraphics[width=0.43\hsize]{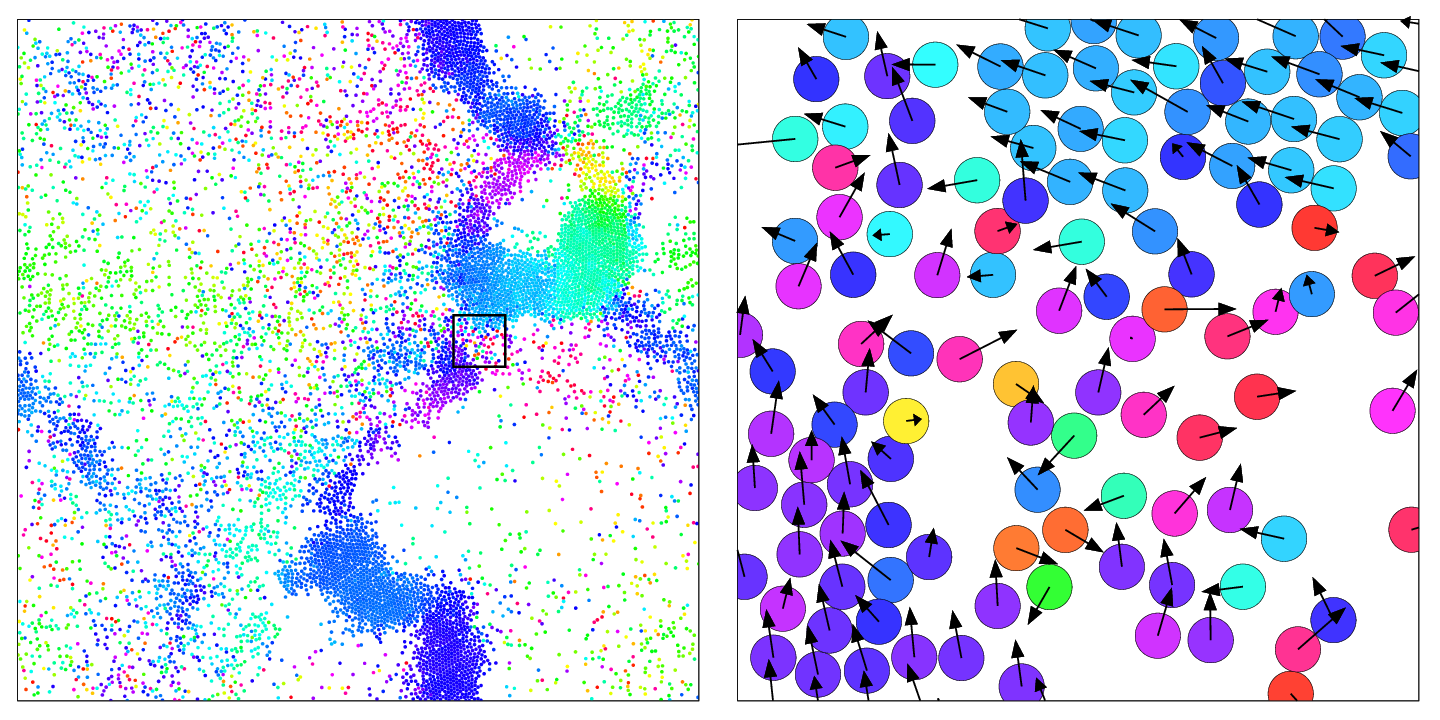}
	\label{fig:4120}
}
\subfloat[$t = 4800$] {
	\centering
	\includegraphics[width=0.43\hsize]{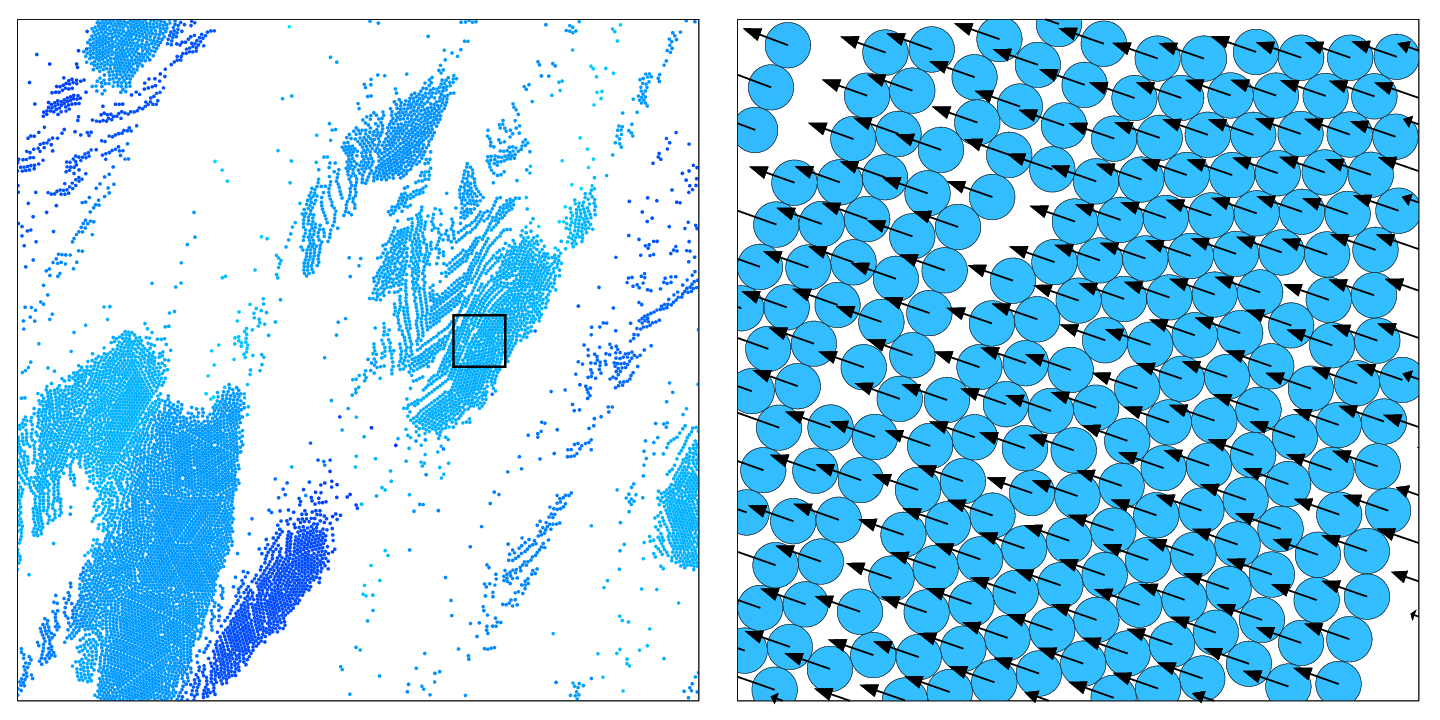}
	\label{fig:4800}
} \\
\subfloat{
	\centering
	\includegraphics[width=0.48\hsize]{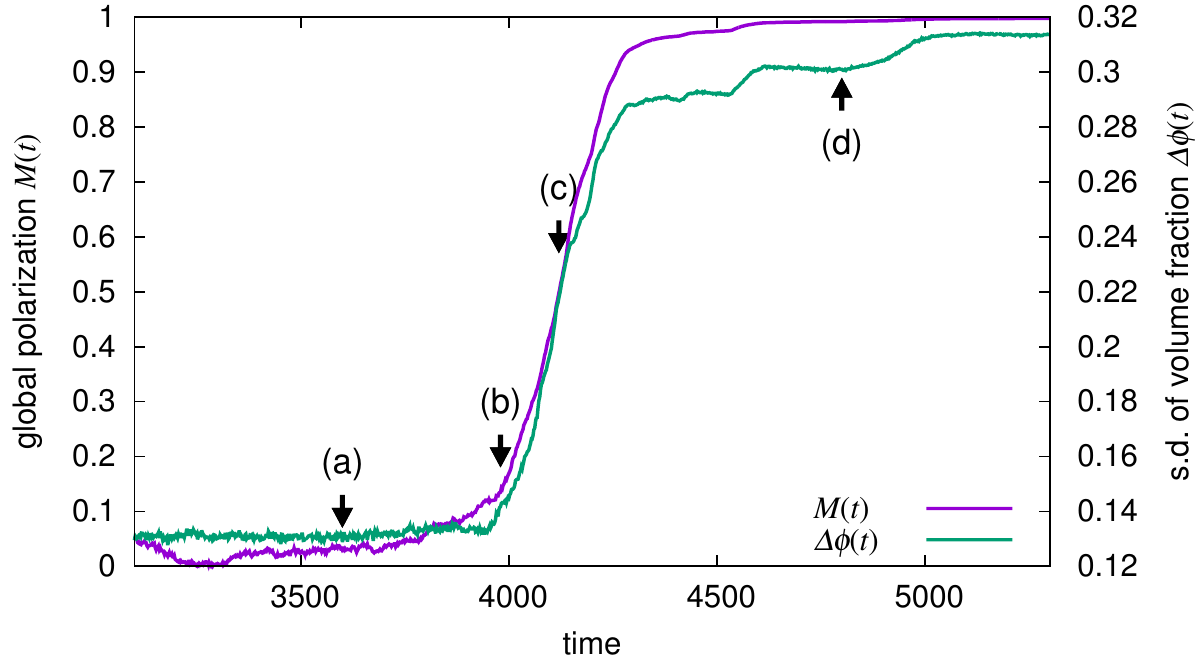}
	\label{timeseries}
}
\caption{Ordering process observed in a representative run for $\Phi = 0.2$, $\gamma = 15.2$, a parameter set which is close to the phase boundary. 
Top: snapshots of the system at different times. At each times, the whole system ($L = 198.2$) is on the left; enlarged image of boxed area of size $15 \times 15$ on the right. The arrows and the colors denote the direction of each particle's velocity. \protect\subref{fig:3600} The system is in isotropic, disordered state. \protect\subref{fig:3980} Dense area appears and local polar order grows within it. \protect\subref{fig:4120} Several locally ordered clusters are formed. \protect\subref{fig:4800} The clusters eventually merge with each other until the whole system moves coherently. 
Bottom: time development of the global polarization $M(t)$ and fluctuation in local volume fraction $\Delta\phi (t)$ from $t=3100$ to $t=5300$. Arrows (a) - (d) correspond to the times in the top figure. }
\label{Snapshots}
\end{figure*}

\section{Model}
We consider $N$ disk particles of equal radius $a$ in a two-dimensional continuous surface. The dynamics is governed by the following deterministic equations \cite{Hiraoka2015SMSEC}: 
\begin{gather}
\frac{\mathrm{d}{\vec{v}_i}}{\mathrm{d}t} = \alpha\hat{\vec{e}}(\psi_i) - \beta \vec{v}_i + \sum_j \vec{f}_{ij}, \label{eom1} \\
\frac{\mathrm{d} \psi_i}{\mathrm{d}t} = \gamma (\theta_i - \psi_i). \label{eom2}
\end{gather}
Eq.~\eqref{eom1} describes the Newtonian equation of motion with velocity $\vec{v}_i = \mathrm{d}\vec{r}_i / \mathrm{d} t$.  The first term of the rhs is the self-propelling force of fixed magnitude $\alpha$ along the direction of ``polarity", an internal degree of freedom defined by an unit vector $\hat{\vec{e}}(\psi_i) \equiv \cos \psi_i \hat{\vec{x}} + \sin \psi_i \hat{\vec{y}}$.  The second term is dissipation proportional to $\vec{v}$, whose strength is determined by $\beta$.  The interaction force is given by binary, short-ranged repulsion: here we assume linear Hookean contact, $\vec{f}_{ij} =-k \left(2a - r_{ij} \right) (\vec{r}_i - \vec{r}_j) / r_{ij}$ if $r_{ij} \equiv | \vec{r}_i - \vec{r}_j | < 2a$ (in contact) and $\vec{f}_{ij} =0$ otherwise.  Eq.~\eqref{eom2} describes the time evolution of the polarity $\psi_i$. When $\psi_i$ deviates from the direction $\theta_i$ of the velocity, it is rotated by a torque proportional to the deviation $\theta_i - \psi_i$ with a coefficient $\gamma$. Thus the equations include damping term for both translational and rotational degrees of freedom. The former is underdamped, while the latter is overdamped.

Note that each parameter gives a different characteristic time: $\tau_\alpha = 2 a \beta \alpha^{-1}$ is the time scale that a particle at the stationary speed takes to run its own diameter; $\tau_\beta = \beta^{-1}$ is the relaxation time of speed; $\tau_k = k^{-1/2}$ is the characteristic time during which two colliding particles are in contact; $\tau_\gamma = \gamma^{-1}$ is the relaxation time of polarity. Without loss of generality, we set unit of length and time as $2a = 1$ and $\beta^{-1} = 1$, and obtain rescaled equations.

\section{Numerical results}
\subsection{Ordering behavior}
We perform particle dynamics simulations of a $N = 10,000$ system in a square box of size $L \times L$ with periodic boundaries. Initial positions and polarities are randomly assigned; overlaps between particles are reduced prior to each run. Unless otherwise specified, we discuss results for $\alpha =1$ and $k = 100$. Under such choice of parameter values, the elasticity is large enough to avoid unphysical situations where particles in contact pass through each other.
A fourth-order Runge-Kutta method is employed for the numerical integration. The time step size is chosen to be sufficiently small compared to both $\tau_k$ and $\tau_\gamma$, either of which defines the shortest time scale of the dynamics in the system.

As shown in Fig.~\ref{Snapshots}, the system exhibits polar ordering and large density fluctuation, the two characteristics also seen in the Vicsek model. The polar order is characterized by the average normalized velocity,
\begin{equation}
\vec{M} =\frac{1}{N} \sum_{i=1}^{N}\vec{\hat{e}}(\theta_i).
\label{global_pol}
\end{equation}
If the system is in random state $M = |\vec{M}| \simeq 0$, while $M = 1$ for a perfectly ordered state.
In order to quantify the density fluctuation, we divide the system into $N_c$ small cells of size $2 \times 2$ and take the standard deviation of the local packing fraction,
\begin{equation}
\Delta \phi = \sqrt{\frac{1}{N_c}\sum_{j=1}^{N_c} \left(\phi_{j}\right)^2 -  \left(\frac{1}{N_c}\sum_{j=1}^{N_c} \phi_{j}\right)^2},
\label{densityfluct}
\end{equation}
where $\phi_{j}$ denotes the local packing fraction in the cell $j$.

Fig.~\ref{Snapshots} shows a typical ordering behavior. Initially, the system is randomized in position and in polarity, so both $M$ and $\Delta \phi$ have small values. When the ordering starts, relatively dense and locally ordered regions appear and grow to form several clusters. The clusters eventually merge with each other until all the particles move into an identical direction. The time series of $M$ and that of $\Delta \phi$ display a simultaneous increase.

\subsection{Phase diagram}
We explore the behavior of the model by implementing a set of simulations with different packing fractions $\Phi = N a^2 \pi / L^2$ and rotational damping parameter $\gamma$. We find that the globally aligned state emerges in the regime where system is dense and rotational damping is weak, while the disordered, isotropic state persists if we set the packing fraction small and the damping parameter large. For $\gamma = 0$, where polarities are never rotated from initial randomized condition, the system maintains a trivial disordered state; however, a small but finite value of $\gamma$ leads to a slow ordering. 

\begin{figure}[tb]
 	\centering
	\includegraphics[width=0.9\hsize]{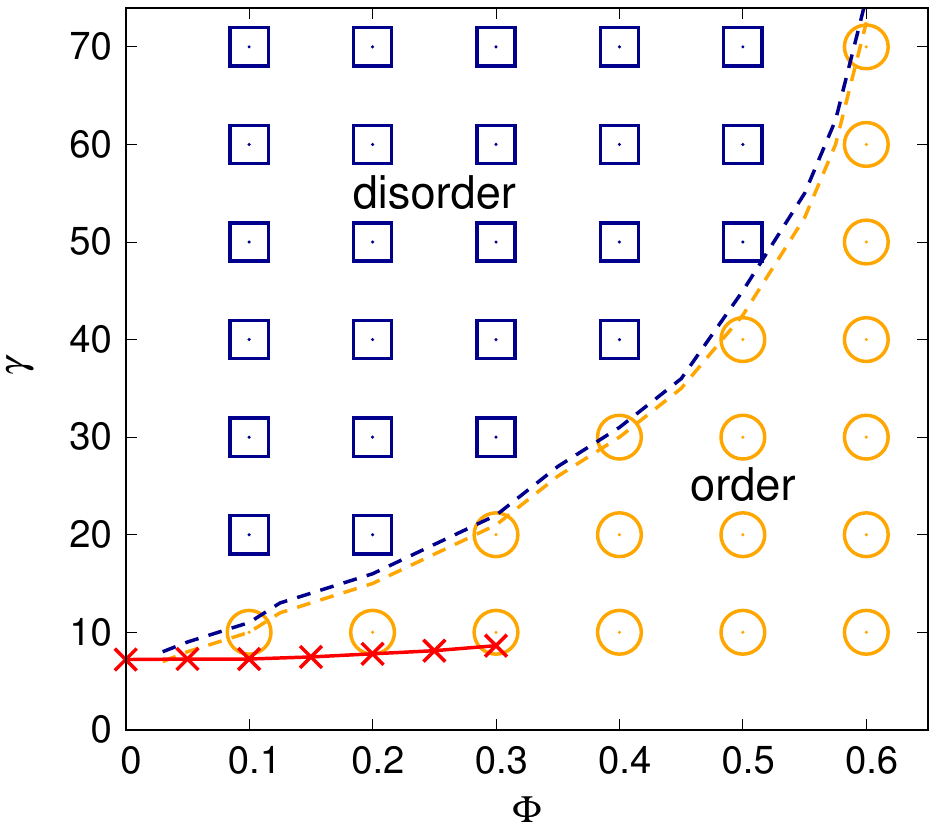}
	\caption{Phase diagram as a function of packing fraction $\Phi$ and damping parameter $\gamma$. Orange circles indicate parameter sets where ordered final state is observed at least for one run; blue squares are where the system remained disordered throughout the simulation time in all the realizations. Orange and blue dashed lines connect the uppermost points in ordered region and the lowermost points in disordered region, respectively, confirmed by a finer parameter search. Red crosses with solid line denote the zero-crossing points from the binary scattering analysis.}
	\label{phasediagram}
\end{figure}

We construct the phase diagram by performing a set of runs (typically 8 to 16) with different initial configurations for a certain simulation time $T$. If polar order, namely $M > 0.8$, is established for one or more runs, the parameter set is classified as a part of ordered region; otherwise, it is in the disordered phase. We choose $N=10,000$ and $T=5,000$ and depict Fig.~\ref{phasediagram}.

In the ordered region near the phase boundary, the system maintains the disordered state (small $M$) until it suddenly transits to the ordered state ($M = 1$). The lifetime $t_w$ of the disordered state, which we refer to as the waiting time, varies depending on the initial configuration. As we increase $\gamma$, $t_w$ tends to be longer and, eventually, ordering behavior does not take place within the simulation time for any realizations.

\begin{figure}[tb]
	\centering
	\includegraphics[width=\hsize]{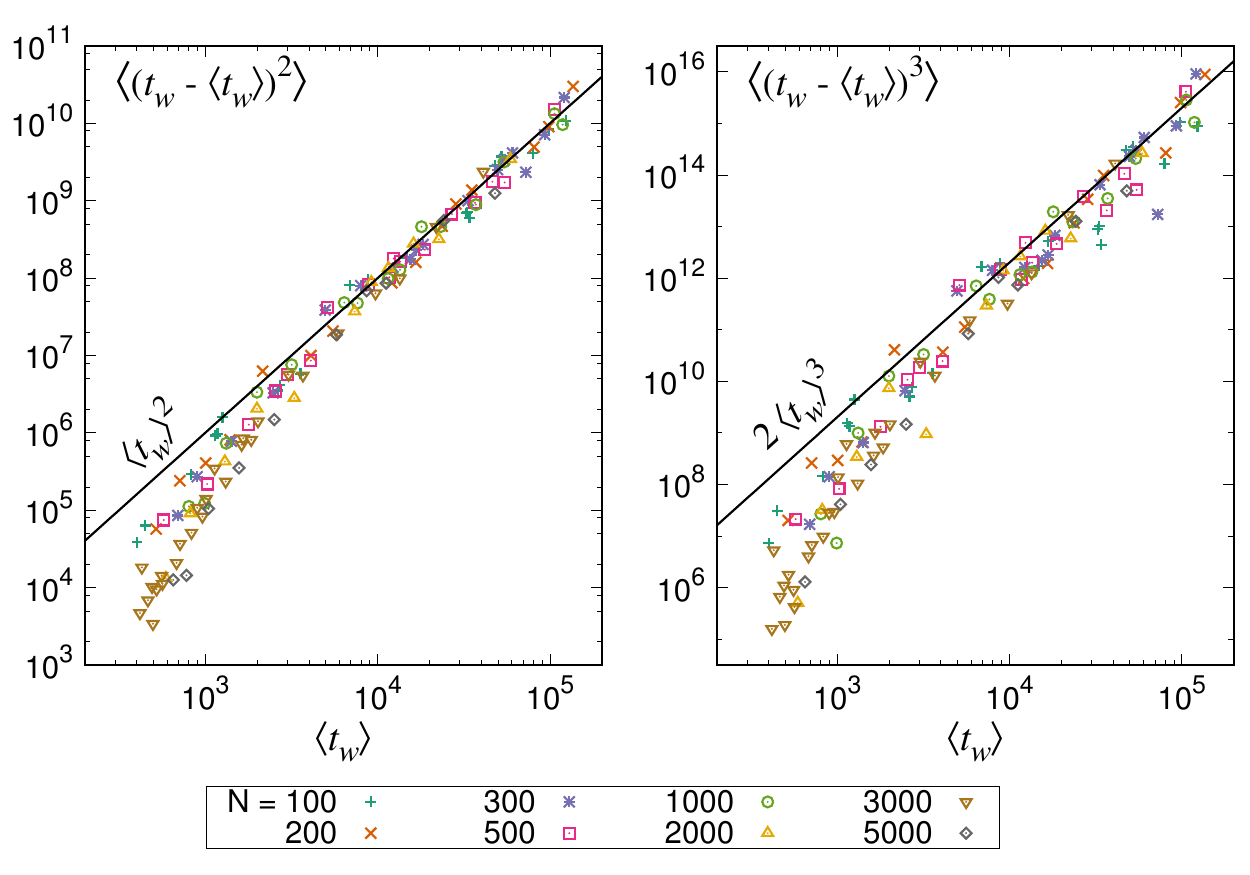}
	\caption {Average waiting time $\langle t_w \rangle$ versus the variance (left) and $\langle t_w \rangle$ versus the third central moment (right). The straight lines represent the expected relations between the moments for exponential distributions.}
	\label{moments}
\end{figure}

We look into the distributions that $t_w$ follows for each parameter set $(N, \Phi, \gamma)$. In Fig.~\ref{moments}, the second central moment $\mu_2 = \langle \left( t_w - \langle t_w \rangle \right)^2 \rangle$ and third central moment $\mu_3 = \langle \left( t_w - \langle t_w \rangle \right)^3 \rangle$ are plotted against the average $\langle t_w \rangle$. In the large waiting time regime, namely where $\langle t_w \rangle > 5,000$, two moments satisfy $\mu_2 = \langle t_w \rangle^2$ and $\mu_3 = 2 \langle t_w \rangle^3$, which is expected for exponential distribution. The fact that the waiting time follows an exponential distribution implies that ordering events occur as Poisson processes involving nucleation phenomena.

\begin{figure}[tb]
	\centering
	\includegraphics[width=\hsize]{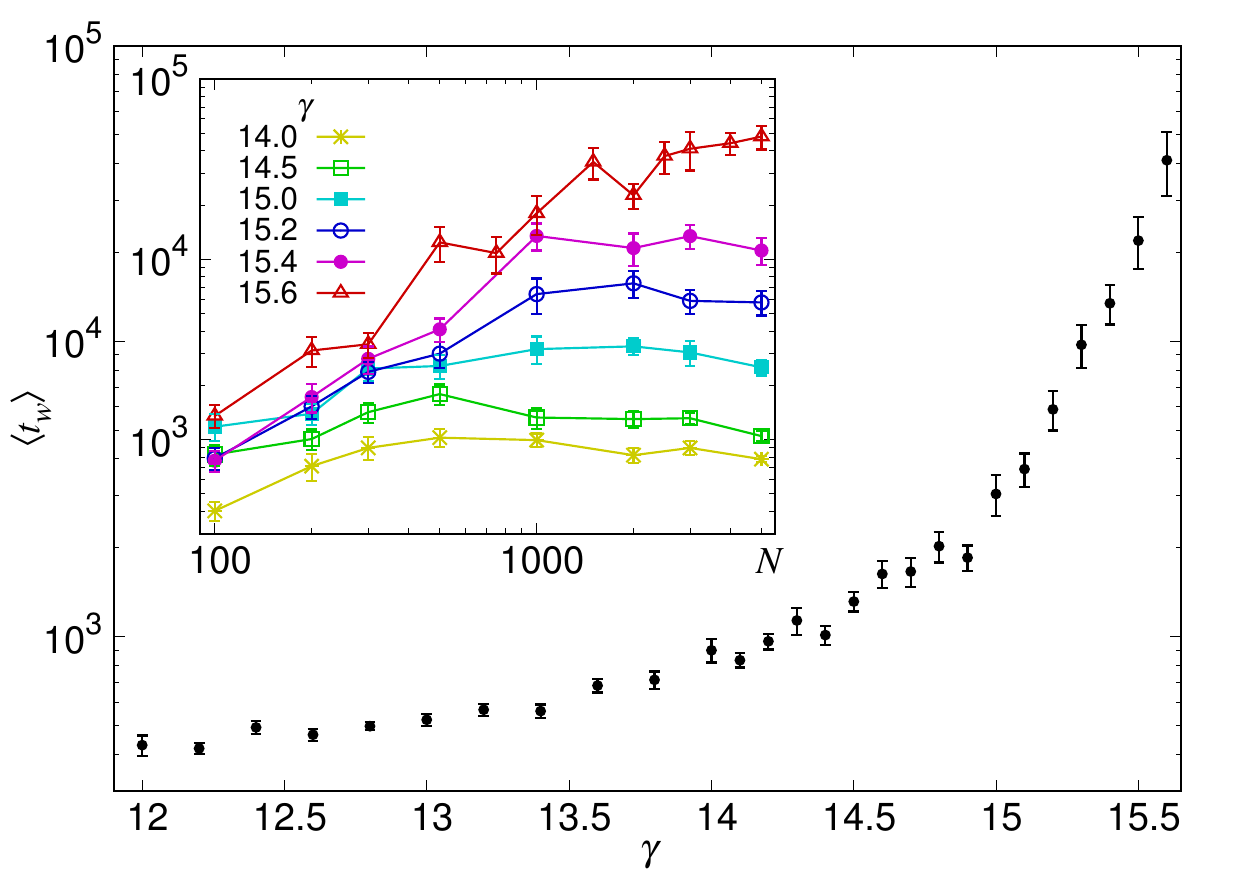}
	\caption {$\langle t_w \rangle$ for system of volume fraction $\Phi=0.2$. The error bars represents the standard error. 
	Main figure: increase as a function of rotational damping parameter $\gamma$. The number of particles $N = 3,000$. The onset of rapid increase is prominent around $\gamma \sim 14.5$. 
	Inset: system size dependence for different values of $\gamma$.}
	\label{waittime}
\end{figure}

Next, we discuss how the phase diagram would be changed if we choose other system sizes and time scales. Finite-size effects are investigated by changing the number of particles $N$ while keeping the volume fraction $\Phi$ fixed. In small systems, $\langle t_w \rangle$ strongly depends on $N$; in larger systems, however, the increase becomes insignificant (Fig.~\ref{waittime}, inset). We confirm that $N = 3,000$ is sufficiently large to avoid the finite-size effects for the time scale that we deal with. 

$\langle t_w \rangle$ displays a rapid increase as a function of $\gamma$ (Fig.~\ref{waittime}). It is difficult to identify the function that fits the growth, but the phenomenological double-well potential picture, which will be described in the next subsection, implies that ordering behavior can occur in any parameter region if we wait long enough. If this is the case, the phase classification inevitably depends on the observation time. Fortunately, the rapid increase, which seems to be exponential or faster, also ensures that the results with an moderately long simulation time provide a good approximation of the results with longer time scales. In other words, the phase diagram would be changed only slightly even if an order of magnitude longer observation time is employed.

\subsection{Dynamics of order parameter}

In this subsection, we show that the dynamics of the order parameter is described as a motion in a double-well potential. This phenomenological picture is consistent with the Poissonian character of the ordering process. It also provides a possible explanation for the divergence of the waiting time: In the ordered regime far from the boundary, the system quickly evolves from an unstable disordered state to an ordered state; near the boundary, however, the disordered state becomes metastable and a nucleation process is necessary to escape from it.

\begin{figure}[tb]
 	\centering
	\includegraphics[width=\hsize]{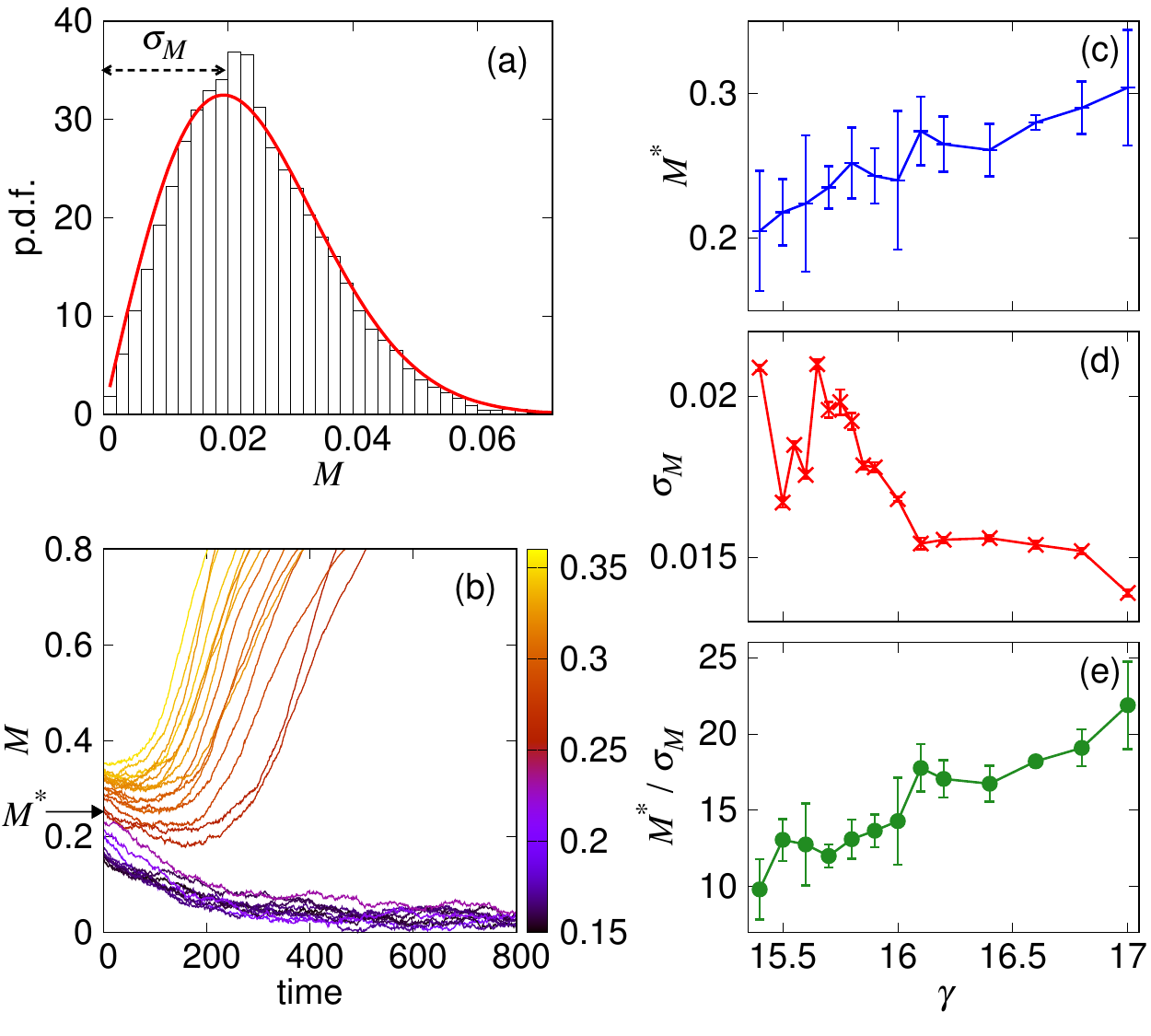}
	\caption{All the figures represent the data for $N = 10,000, \Phi = 0.2$. 
	(a) Normalized distribution of $M$ in a disordered state, obtained from four independent runs with $\gamma = 15.8$ and $T = 5,000$, fitted by probability density function of a Rayleigh distribution with scale parameter $\sigma_M$ (red solid curve). 
	(b) Time series of $M$ in initially asymmetric runs with $\gamma = 15.8$. Colors correspond to the initial values $M(0)$ as indicated by the color bar. Polar order quickly emerges with $M(0) > M^*$. 
	(c)(d) $\gamma$ dependence of  $M^*$, which decreases as $\gamma$ approaches $\gamma_c \simeq 15.6$ from above, and of $\sigma_M$, which increases, respectively.
	(e) The ratio between the two values decreases, suggesting that the fluctuation of $M$ is more likely to lead to an escape from the metastable, disordered state by crossing the potential barrier.}
	\label{flucasym}
\end{figure}

When the system is in the disordered state, the global polarization vector $\vec{M}$ fluctuates around $\vec{0}$. The microscopic origin of the fluctuation is change in polarity of the particles caused by collisions to their neighbors. We expect that the collision events are uncorrelated to each other, and the time development of $\vec{M}$ can be treated as Brownian motion in the potential. Assuming that the potential is harmonic around $\vec{M}=\vec{0}$, we expect that $\vec{M}(t)$ constitutes a two-dimensional Ornstein-Uhlenbeck process\cite{Unlenbeck1930}. The stationary distribution of the radial component $M(t)$ should be given by a Rayleigh distribution, 
\begin{equation}
f(M) = \frac{M }{\sigma_{M}^2} \exp \left(-\frac{M^2}{2 \sigma_{M}^2} \right),
\end{equation}
where the scale parameter $\sigma_{M}$ is the characteristic amplitude of the fluctuation. In fact, the observed distribution can be fitted reasonably by a Rayleigh distribution (\mbox{Fig.~\ref{flucasym}(a)}), providing an estimate of $\sigma_M$ as
\begin{equation}
\sigma_{M}^2 = \frac{1}{2t}\int_{0}^{t} \left( M(t') \right)^2 dt'.
\end{equation}

Aside from the fluctuation amplitude, the escape rate is determined by the ``position" of the barrier in order parameter space, which can be estimated by preparing a system in which a fraction of particles are given the same initial polarity so that initial polarization $M(0)$ has a finite value. The initial positions are uniformly distributed both for aligned and unaligned particles. The double-well potential picture suggests that the time development of the system depends on which side of the barrier the initial state is situated: Systems with $M(0)$ smaller than a certain value relax to the disordered state and those with $M(0)$ larger than the same threshold goes to the ordered state. The threshold $M^*$ indicates the position of the barrier (\mbox{Fig.~\ref{flucasym}(b)}).

%%%%A quick evolution into a polarized state implies that the initial state is already on the polarized side of the barrier; if the system is pulled back to a disordered state, it is situated nearer the disordered state (\mbox{Fig.~\ref{flucasym}(b)}). The threshold value $M^*$ indicates the position of the barrier in the order parameter space. 

Decreasing $\gamma$ in the disorder region towards the transition point, the threshold $M^*$ decreases (\mbox{Fig.~\ref{flucasym}(c)}) and the fluctuation amplitude $\sigma_M$ increases (\mbox{Fig.~\ref{flucasym}(d)}). These results imply that the escape rate increases and the average time until spontaneous polarization occurs will be shorter.

Due to the absence of noise, a fully-ordered system does not evolve back to a disordered state; the ordered state is an absorbing state. 
We also study the stability of the ordered state numerically. The initial polarities are set to be aligned except for a small fraction ($5$ to $20\%$ of all particles)  with random directions. The initial positions are, again, uniformly distributed both for aligned and unaligned particles. In all the realizations, the system quickly relaxes to a fully ordered state, even if we choose a parameter set deep in the disordered phase, such as $\Phi = 0.2, \gamma = 50$. This result suggests that the ordered state is stable against perturbation.

\section{Binary scattering at dilute limit}
In this section, we give a simple explanation to understand the mechanism that underlies the ordering behavior shown in the previous section by focusing on the binary particle collision process, a method also employed in \cite{Hanke2013}. Here we assume that the system is dilute ($\Phi \to 0$) so that the collisions are uncorrelated with each other, and that both the velocity and the polarity are fully relaxed before every collision.

\begin{figure}[tb]
	\centering
	\includegraphics[width=0.95\hsize]{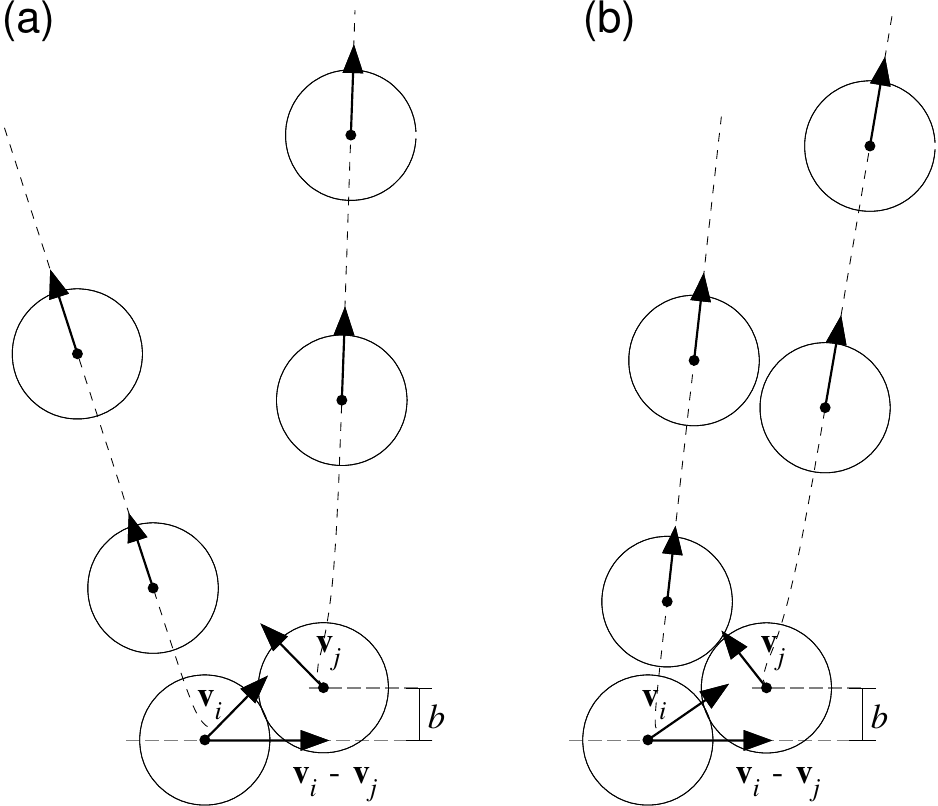}
	\caption{Illustration of binary scattering. 
	(a) In dilute limit, the velocity of the two particles are fully relaxed ($v_i = v_j = \alpha$) before collision. The geometry of binary collision is fully specified by the relative angle $\theta_{ij} = \arccos \left( \vec{v}_i \cdot \vec{v}_j / v_i v_j\right)$ of the velocities and impact parameter $b$. 
	(b) For finite densities, the velocities may not be at the stationary speed at the moment of contact. Here, the relative velocity $\vec{v}_i - \vec{v}_j$ and the impact parameter $b$ are equal to those in (a), but the velocities are set to be $v_i = 1.1 \alpha, v_j = 0.8 \alpha$. The consequent trajectory differs from the dilute case.}
	\label{collisiongeo}
\end{figure}

Let us consider a binary scattering process between particle $i$ and $j$ (Fig.~\ref{collisiongeo}). Since we assume the rotational invariance, the geometry of the moment of contact is fully specified by two scalar parameters: the impact parameter $b =  \sqrt{r_{ij}^2 - \vec{r}_{ij} \cdot (\vec{v}_i - \vec{v}_j) / v_{ij}} \in [0, a)$, where $v_{ij} = {|\vec{v}_i - \vec{v}_j|}$, and the relative angle $\theta_{ij}  = {|\theta_i - \theta_j|} \in (0, \pi]$. The impact parameter shows the perpendicular offset of the two bodies' center of mass from head on collision. If $b = 0$ the collision is head on whereas it is a miss if $b > a$.

Instantaneous alignment of the two particles is characterized by two-particle polarization,
\begin{equation}
M^{(2)} =\frac{1}{2} \left| \hat{\vec{e}}(\theta_i) + \hat{\vec{e}}(\theta_j) \right|,
\label{eq:2polarization}
\end{equation}
which corresponds to Eq.~(\ref{global_pol}) with $N=2$. 
We measure the post-collisional two-particle polarization $M^{(2)}_{\mathrm{out}}$ at a point where the polarities and the velocities are fully relaxed, and compare it to the pre-collisional polarization $M^{(2)}_{\mathrm{in}}$. The increment $\Delta M^{(2)} =M^{(2)}_{\mathrm{out}} - M^{(2)}_{\mathrm{in}}$ indicates the magnitude of parallel alignment caused by the scattering process. 

Assuming the system is homogenous and isotropic, two particles should collide in the relative angle of $\theta_{ij}$ with a probability proportional to the relative velocity $v_{ij}$; the impact parameter $b$ should be uniformly distributed. 
The average tendency of binary alignment, as a function of $\gamma$, is then obtained by taking an integral weighted by the ``scattering cross section,"
\begin{equation}
\langle \Delta M^{(2)} \rangle = \frac{1}{C} \int_0^\pi \int_0^{2a}  \left| \sin \left( \frac{\theta_{ij}}{2}\right)\right|\, \Delta M^{(2)}(\theta_{ij}, b) \, \mathrm{d}b \, \mathrm{d} \theta_{ij} ,
\label{scattcross}
\end{equation}
where $C$ is a normalization constant.

\begin{figure}[tb]
	\centering
	\includegraphics[width=\hsize]{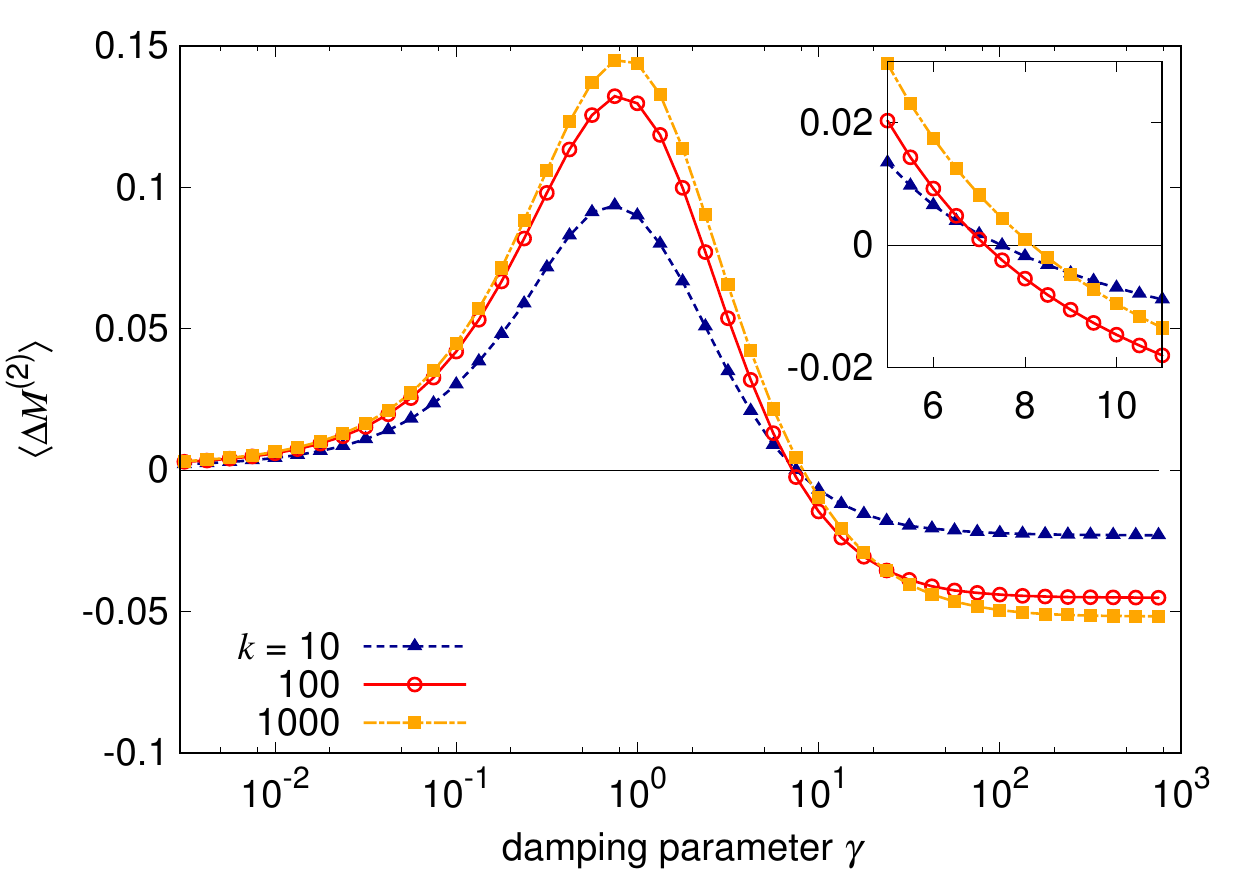}
	\caption{Average binary alignment $\langle \Delta M^{(2)} \rangle$ as a function of angular damping parameter $\gamma$. The tendency of alignment reaches its maximum at $\gamma  \sim 1$ and decreases to take negative values for larger $\gamma$. Inset: zoom of the zero-crossing point.}
	\label{binary_dilute}
\end{figure}

The result shown in Fig.~\ref{binary_dilute} indicates that the alignment tendency has a maximum at $\gamma \sim 1$. For $\gamma \to 0$, which corresponds to the regime where angular relaxation is slow, $\langle \Delta M^{(2)} \rangle$ goes to zero. For large $\gamma$, namely $\gamma \to \infty$, $\langle \Delta M^{(2)} \rangle$ has a negative value.

Qualitative explanation is as follows. If the rotational damping is weak, the polarity of two particles remain unchanged, so the directions of motion will eventually restored to the original direction. In contrast, if the damping is strong, the polarity rotates itself quickly to follow the change in the direction of motion. It is with an intermediate value of the damping strength that the motion of two bodies align parallel, due to the competing effect of repulsive collision and subsequent rotational damping.

This argument is consistent with the results obtained from the many-particle simulations: First, the ordering in many-body system is the fastest in the parameter region that maximizes the value of $\langle \Delta M^{(2)} \rangle$; second, the transition point in a dilute system is in agreement with the zero-crossing point (Fig.~\ref{phasediagram}). These agreements imply that the onset of the collective motion arises from iteration of binary scattering, although one will have to take into account many-body correlation for the late stage of the ordering process, where the isotropic and uncorrelated conditions no longer hold. 

\section{Binary scattering at finite densities}
The analysis in the previous section is only valid for dilute limit, where the velocity and the polarity of the particles are fully relaxed before every collisional event. At finite densities, however, the relaxation may not complete between collisions (Fig.~\ref{collisiongeo}).

\begin{figure}[tb]
 	\centering
	\includegraphics[width=0.95\hsize]{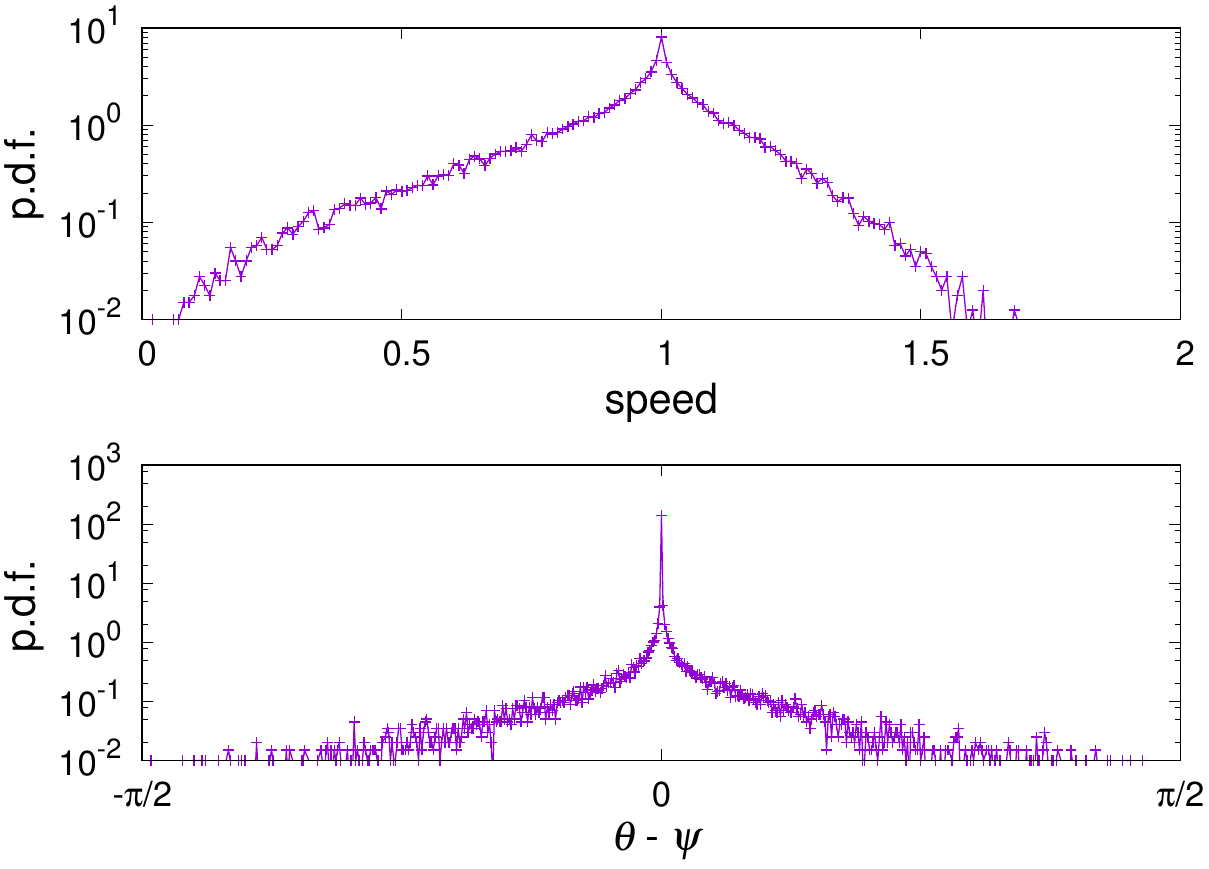}
	\caption{Instantaneous distribution of speed (top) and of angular deviation between velocity and polarity (bottom) in a run with parameter values in the disordered phase ($\Phi = 0.2, \gamma = 18, t = 5,000$). The speed is distributed broadly, while the angle distribution has a much narrower peak at $0$, suggesting most of the particles are relaxed in terms of rotational degree of freedom.}
	\label{dist}
\end{figure}

For the parameter values we choose ($\alpha = 1$ and $k = 100$), $\gamma \gg \beta = 1$ in the disordered phase, meaning that relaxation of polarity is much faster than that of velocity. We check this by measuring the speed $v_i = |\vec{v}_i|$ and the deviation $\theta_i - \psi_i$ between direction of velocity and polarity of all the particles directly from the many-particle simulation. Fig.~\ref{dist} shows that the distribution of $v_i$ is considerably broad compared to the polarity distribution, which has a narrow peak at the point where $\theta_i = \psi_i$. We will therefore take into account the speed distribution while safely neglecting the angular deviation in the following.

We apply a similar binary scattering study as in the previous section, except that $M^{(2)}_{\mathrm{out}}$ is not defined at the point where the relaxation is complete, but at the point where the particle reaches the distance of the mean free path (mfp) away from the binary contact.

To calculate the mfp, we assume that every particle moves at speed $v$. The magnitude of relative velocity between two particles is  
\begin{equation}
v_r = 2v \sin \left(\frac{\theta_{ij}}{2}\right).
\end{equation}
If collisions are uncorrelated to each other, the mean relative velocity is
\begin{equation}
\langle v_r \rangle = \frac{1}{\pi} \int_0^\pi 2v \sin \left(\frac{\theta_{ij}}{2}\right)\, \mathrm{d} \theta_{ij} = \frac{4v}{\pi}.
\end{equation}
Since the scattering cross section $s =4a$ and the number density $n = \Phi / a^2 \pi$, the mfp is given as 
\begin{equation}
\lambda = \frac{v}{sn\langle v_r \rangle} = \frac{a \pi^2}{16 \Phi}.
\end{equation}

Suppose particles $i$ and $j$ of speed $v_i$ and $v_j$ come into contact and are scattered. We assume that the speed of the particles independently obeys an identical distribution $f(v)$. The speeds evolve to $v_i'$ and $v_j'$ when the particles travels the distance of mfp after the collision. The post-collision speed distribution can be written as
\begin{equation}
f(v') = \hat{S}f(v), 
\end{equation}
where $\hat{S}$ denotes the scattering operator.

If the system is in the disordered phase, the distribution is stationary, so the following self-consistent condition should be satisfied:
\begin{equation}
\hat{S}f_s(v)= f_s(v),
\end{equation}
where $f_s(v)$ is the stationary distribution. According to the Perron-Frobenius theorem\cite{Berman1979}, $f_s(v)$ is the eigenfunction of operator $\hat{S}$ which corresponds to the eigenvalue one. 

$\hat{S}$ is numerically derived by mapping pre-collision speeds $(v_i, v_j)$ onto post-collision speeds $(v'_i, v'_j)$. We divide the $v$-space into bins and simulate a binary scattering process for the representative value for each bin to obtain the matrix $S$. Applying the power iteration method, we yield $f_s(v)$ as the eigenvector of $S$.

Once $f_s(v)$ is known, the averaged increment in the two particle polarization is calculated as 
\begin{multline}
\langle \Delta M^{(2)} \rangle = \frac{1}{C} \int_{0}^{\infty}\mathrm{d}v \int_{v_i - v_j}^{v_i + v_j} \mathrm{d}v_{ij} \int_{-a}^{a} \mathrm{d}b \\
\frac{f_s(v)}{v_{ij}} \Delta M^{(2)}(v_i, v_j, v_{ij}, b),
\end{multline}
%\begin{equation}
%\langle \Delta M^{(2)} \rangle = \frac{1}{C} \int_{0}^{\infty}\mathrm{d}v \int_{v_i - v_j}^{v_i + v_j} \mathrm{d}v_{ij} \int_{-a}^{a} \mathrm{d}b \, \frac{f_s(v)}{v_{ij}} \Delta M^{(2)}(v_i, v_j, v_{ij}, b),
%\end{equation}
where $v_{12} = |\vec{v}_1 - \vec{v}_2|$ and $C$ is a normalization constant.

Again, by numerically calculating the point where $\langle \Delta M^{(2)} \rangle$ crosses zero, we obtain the estimated phase boundary. However, the estimation deviates from the results of many-body simulations, as shown in Fig.~\ref{phasediagram}. The binary scattering approach is based on three assumptions: (i) the collisions are uncorrelated to each other; (ii) the motion of the particles is isotropic; (iii) the system is homogeneous, i.e., the density does not fluctuate. We suppose that these assumptions basically contribute to suppress the emergence of ordering behavior and that the deviation with the results from many-body simulations stems from the failure of one or more of the assumptions. A more detailed analysis of this point, particularly a closer look into the local quantities, is left for future work.

\section{Summary}
In this work we consider a simple model of repulsive self-propelled particle systems. Numerical simulations indicate that the system exhibits polar-ordering and a transition between ordered phase and disordered phase; the phase diagram is depicted as a function of the packing fraction and the rotational damping strength. We show that the slowdown of the dynamics at the proximity of the phase boundary is caused by metastability of the disordered state, and that transition to a polarized state requires a nucleation process. The physics that induces the emergence of the collective motion is the iteration of binary collision and the rotational relaxation afterwards. This argument is supported by the fact that the binary scattering analysis predicts the transition point from the many-particle simulation in the dilute limit. 
For finite densities, the approach deviates from the actual boundary because the disordered state is unstable against the many-particle correlations, anisotropy of the collisions, and the density fluctuation in the system. Recently, much effort has been devoted to derive a hydrodynamic description directly from the microscopic dynamics of SPP\cite{Bertin2006, Baskaran2008, Peshkov2014, Ihle2014}. Applying the kinetic approaches to the model we present will be an interesting direction of future work.

This model is consistent with some other models presented in previous studies. 
(i) Soft particle models for epithelial cell migration developed in \cite{Szabo2006, Henkes2011} has a first order equation of motion instead of Eq.~\eqref{eom1}, namely $\dot{\vec{r}_i} = v_0\hat{\vec{e}}_{\psi_i} + \sum_j \vec{f}_{ij}$, where $v_0$ is a constant speed. This model is  equivalent to ours when we take the overdamped limit, $\beta \to \infty$ with $\alpha / \beta = \mathrm{const}$. 
(ii) The model mentioned in \cite{Vicsek2012} and explored in \cite{Hanke2013} does not have polarity degree of freedom and employs the equation of motion as follows: $\dot{\vec{v}_i} = v_0 \hat{\vec{v}}_i - \vec{v}_i + \sum_j \vec{f}_{ij}$,  which means self-propulsion is always directed towards the direction of motion. This model corresponds to another limit in our model, $\gamma \to \infty$. 
(iii) The social force model for pedestrian movement \cite{Helbing2000} is given by following dynamics: $\dot{\vec{v}_i} = v_0 \hat{\vec{e}}_i^0 - \vec{v}_i +  \sum_j \vec{f}_{ij}$, where $\hat{\vec{e}}_i^0$ denotes the individual's desired direction. This equation of motion is analogous to $\gamma = 0$ case in our model, where polarity of particles are never changed, although the particular form of the short-range repulsive interaction is different. 
Hence, our model can be regarded as a generalized model that bridges the above three models. While these previous models incorporate noise as a control parameter, we show that the phase transition is realized even in the absence of noise. In conclusion, our findings elucidate the universality of collective ordering behavior in repulsive SPP systems.

\section*{Acknowledgments}
T. H. thanks Koji Oishi for helpful discussions. The authors also appreciate comments and suggestions from anonymous reviewers, which contributed greatly to improve the paper. This work was supported by CREST program of Japan Science and Technology Agency.

\bibliography{repulsiveSPP}

\end{document}